# Dust in a merging galaxy sequence: the SCUBA view

A. Georgakakis, *, E. M. Xilouris, *, A. Misiriotis, † and V. Charmandaris, **

*Institute of Astronomy & Astrophysics, National Observatory of Athens, I. Metaxa & Vas. Pavlou, Palaia Penteli, 15236, Athens, Greece
†University of Crete, Physics Department, PO Box 2208, 71003 Heraklion, Crete, Greece
**Astronomy Department, Cornell University, Ithaca, NY 14853, USA

**Abstract.** We investigate the cold and warm dust properties during galaxy interactions using a merging galaxy sample ordered into a chronological sequence from pre- to post-mergers. Our sample comprises a total of 29 merging systems selected to have far-infrared and sub-millimeter observations. We use the 100–to–850$\mu$m flux density ratio, $f_{100}/f_{850}$, as a proxy to the mass fraction of the warm and the cold dust in these systems. We find evidence for an increase in $f_{100}/f_{850}$ along the merging sequence from early to advanced mergers and interpret this trend as an increase of the warm relative to the cold dust mass. We argue that the two key parameters affecting the $f_{100}/f_{850}$ flux ratio is the star-formation rate and the dust content of individual systems relative to the stars. Using a sophisticated model for the absorption and re-emission of the stellar UV radiation by dust we show that these parameters can indeed explain both the increase and the observed scatter in the $f_{100}/f_{850}$ along the merging galaxy sequence.

## INTRODUCTION

One of the revolutionary results from the SCUBA bolometer array on the JCMT has been the direct detection at submillimeter (submm) wavelengths ($450,850\mu m$) of cold dust (T=15-25 K) in isolated quiescent spiral galaxies that dominates the overall dust content. This component remained undetected by IRAS (60, 100 $\mu m$) that probes warmer dust (T$>$ 30K) and is only detectable at longer submm wavelengths.

In more active interacting systems submm observations also suggest the presence of cold cirrus. Unlike quiescent spirals however, mergers also show a prominent warm dust component (T=30-60 K) most likely heated by the enhanced star-formation.

In this study we explore the role of galaxy interactions in heating the dust and hence, modifying the relative amounts of the cold and the warm dust component in galaxies. Our interacting galaxy sample comprises well separated spirals, systems close to nuclear coalescence and young merger remnants.

## THE SAMPLE

The interacting galaxy sample used here is compiled from the Georgakakis et al. [1] study and comprises disk galaxy mergers of similar mass spanning a wide range of interactions stages: well separated spirals, systems close to nuclear coalescence and young merger remnants. All the systems employed here have submm data available from the SCUBA archive.

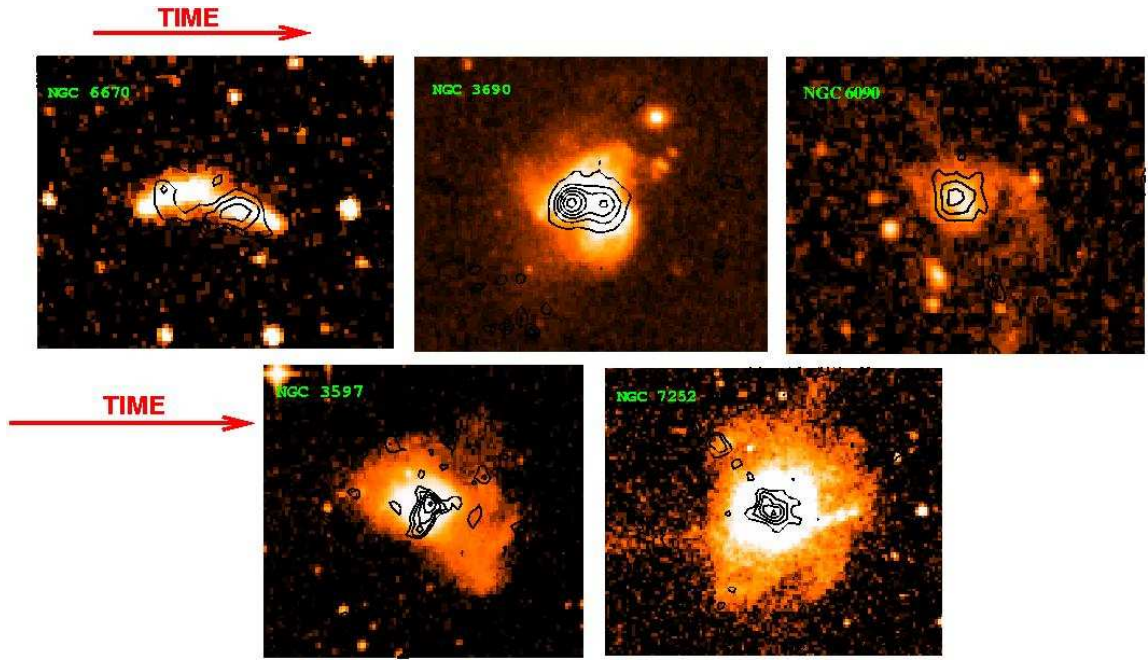

**FIGURE 1.** An example of merging galaxy sequence comprising well separated interacting galaxies on the left, progressively more advanced mergers (from left to right) and mergers remnants (bottom right). The merging evolves from left to right. The contours are submm 850 $\mu m$ data plotted over the optical DSS images.

To explore the evolution of the galaxy dust properties at different times during the interaction we order our merging systems into a chronological sequence. For this we use the 'age' parameter providing an estimate of the evolutionary stage of each system relative to the time of the merging of the two nuclei. Negative 'ages' are for pre-mergers while positive 'ages' correspond to merger remnants. An example of a merging sequence is shown in Figure 1.

For pre-mergers the 'age' is estimated by dividing the projected separation of the two nuclei by an (arbitrary) orbital decay velocity v=30km/s. For post-mergers we adopt the evolutionary sequence defined by Keel & Wu [2] using dynamical and morphological criteria. It should be stressed that the 'age' parameter for both pre- and post-mergers does not represent an absolute galaxy age but is an indicator of the evolutionary stage of the interaction.

## RESULTS

We quantify the dust properties of our sample using the 100–to–850 $\mu m$ flux density ratio, $f_{100}/f_{850}$. This provides an estimate of the relative mass fraction of the warm and the cold dust component. The submm is sensitive to cold dust (T=15-25 K) with the far-infrared (FIR) probing the warmer dust (T> 30K).

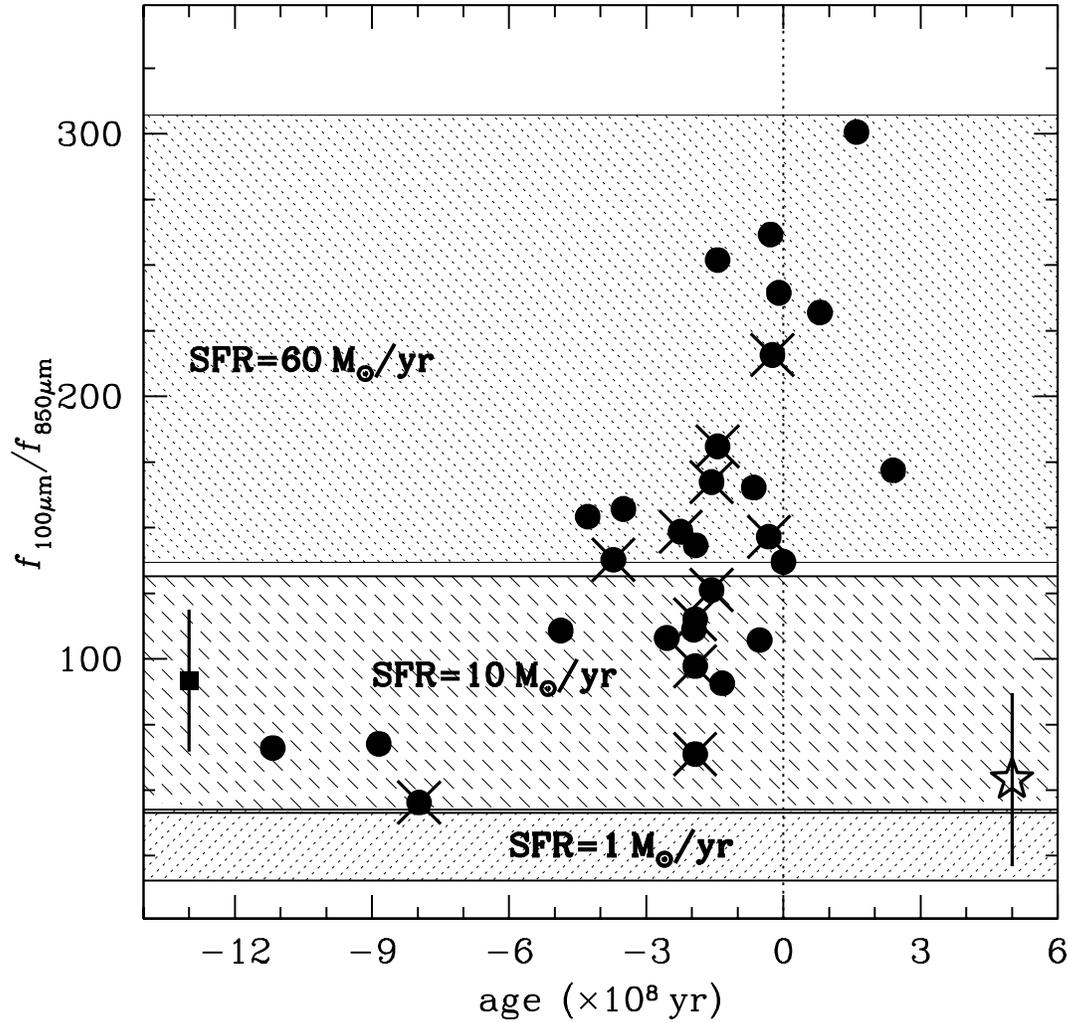

**FIGURE 2.** $f_{100}/f_{850}$ flux ratio against the age parameter for our sample of interacting galaxies. The filled circles are the interacting galaxy sample in this study. A cross on top of symbol signifies an AGN in the catalogue of Veron-Cetty & Veron [4]. The square corresponds to the mean $f_{100}/f_{850}$ of isolated spirals from the sample of Misiriotis et al. [3] arbitrarily plotted at an age parameter of −13. The star is the mean $f_{100}/f_{850}$ of ellipticals from the sample of Temi et al. [5] arbitrarily plotted at an age parameter of +5.

Figure 2 plots the ratio $f_{100}/f_{850}$ as a function of the age parameter for our interacting galaxy sample. There is evidence for an increase of $f_{100}/f_{850}$ as the interaction progresses from well separated spirals toward the final stages of the merger. The Spearman rank test suggests that the correlation is significant at 3.5 $\sigma$ confidence level. This statistically significant correlation is not due to $f_{100}$ or $f_{850}$ being individually correlated with the age parameter.

This trend can be interpreted as an increase of the warm relative to the cold dust mass from isolated spirals and early interacting systems to advanced mergers. Such a transition of the cold dust to warmer temperatures can be attributed to heating by the more intense UV radiation field of the enhanced star-formation triggered by the merging.

We attempt to interpret the trend between $f_{100}/f_{850}$ and age using the model of Misiriotis et al. [3] that describes the absorption of UV/optical photons by dust and their re-emission at FIR and submm wavelengths. This model produces UV to submm SEDs for a given input system geometry, stellar population (e.g. young and old stars), dust content and total galaxy mass.

The dust mass of the fiducial galaxy is assumed to be in the range $10^7 - 10^8$ M$_\odot$, similar to the dust masses of our systems. The stellar content is directly associated with the density of the UV radiation responsible for the heating of the dust and is parameterized by the star-formation rate. Here we use three different values for the star-formation rate SFR=1, 10 and 60 M$_\odot$/yr. For a given SFR and dust mass we derive the $f_{100}/f_{850}$ ratio predicted by the model. The model predictions for different SFRs are shown with the shaded regions in Figure 2. The lower and upper boundaries of these regions correspond to dust masses of $10^8$ and $10^7$ M$_\odot$ respectively.

The model predicts that higher SFRs heat the dust resulting in higher $f_{100}/f_{850}$ ratios. Also, keeping the SFR fixed, more dusty systems have lower $f_{100}/f_{850}$ which can be attributed to more efficient warming of the dust, i.e. higher fraction of UV/optical photons absorbed by the dust, heating it to temperatures T$\leq$ 30K and then being re-emitted primarily at submm wavelengths (850$\mu m$). This may also partly explain the scatter seen in the figure since the relative fraction of the warm–to–cold dust for a given SFR varies substantially with dust content.

## CONCLUSIONS

Using a merging galaxy sequence comprising early interacting spirals, systems close to nuclear coalescence and merger remnants we explore the evolution of dust properties during galaxy mergers. We find evidence for dust heating from well separated young mergers to advanced interacting galaxies. Using a model that describes the absorption of the light by dust and its re-mission to FIR and submm wavelengths we argue that the observed trend can be attributed to the enhanced star-formation rate during interactions as well as differences in the dust masses of individual systems.